\documentclass[twocolumn]{aastex631}

\usepackage{multirow}
\usepackage{graphicx}

\begin{document}

\title{Atmospheric parameters and abundances of cool red giant stars}
\shorttitle{Cool red giant properties}
\shortauthors{Dencs et al.}
\correspondingauthor{Z. Dencs}
\email{dencs.zoltan@csfk.org}

\author{Z. Dencs}
\affiliation{Konkoly Observatory, HUN-REN Research Centre for Astronomy and Earth Sciences, Konkoly Thege Mikl\'os \'ut 15-17, H-1121 Budapest, Hungary}
\affiliation{ELTE E\"otv\"os Lor\'and University, Gothard Astrophysical Observatory, Szent Imre h. u. 112, 9700, Szombathely, Hungary}
\affiliation{MTA-ELTE Lend\"ulet "Momentum" Milky Way Research Group, Hungary}

\author{A. Derekas}
\affiliation{ELTE E\"otv\"os Lor\'and University, Gothard Astrophysical Observatory, Szent Imre h. u. 112, 9700, Szombathely, Hungary}
\affiliation{MTA-ELTE Lend\"ulet "Momentum" Milky Way Research Group, Hungary} 
\affiliation{HUN-REN-SZTE Stellar Astrophysics Research Group, H-6500 Baja, Szegedi \'ut, Kt. 766, Hungary} 

\author{T. Mitnyan} 
\affiliation{HUN-REN-SZTE Stellar Astrophysics Research Group, H-6500 Baja, Szegedi \'ut, Kt. 766, Hungary}
\affiliation{Baja Astronomical Observatory of University of Szeged, H-6500 Baja, Szegedi \'ut, Kt. 766, Hungary} 

\author{M.~F. Andersen}
\affiliation{Department of Physics and Astronomy, Aarhus University. Ny Munkegade 120, 8000 Aarhus C, Denmark}

\author{B. Cseh}
\affiliation{Konkoly Observatory, HUN-REN Research Centre for Astronomy and Earth Sciences, Konkoly Thege Mikl\'os \'ut 15-17, H-1121 Budapest, Hungary}
\affiliation{CSFK, MTA Centre of Excellence, Budapest, Konkoly Thege Mikl\'os \'ut 15-17., H-1121, Hungary}

\author{F. Grundahl}
\affiliation{Department of Physics and Astronomy, Aarhus University. Ny Munkegade 120, 8000 Aarhus C, Denmark}

\author{V. Heged\H{u}s}
\affiliation{MTA-ELTE Lend\"ulet "Momentum" Milky Way Research Group, Hungary}
\affiliation{ELTE E\"otv\"os Lor\'and University, Doctoral School of Physics, P\'azm\'any P\'eter s\'et\'any. 1/A, 1117 Budapest, Hungary}

\author{J. Kov\'acs}
\affiliation{ELTE E\"otv\"os Lor\'and University, Gothard Astrophysical Observatory, Szent Imre h. u. 112, 9700, Szombathely, Hungary}
\affiliation{MTA-ELTE Lend\"ulet "Momentum" Milky Way Research Group, Hungary} 
\affiliation{HUN-REN–ELTE Exoplanet Systems Research Group, Szent Imre h. u. 112., Szombathely H-9700, Hungary}

\author{L. Kriskovics} 
\affiliation{Konkoly Observatory, HUN-REN Research Centre for Astronomy and Earth Sciences, Konkoly Thege Mikl\'os \'ut 15-17, H-1121 Budapest, Hungary} 
\affiliation{CSFK, MTA Centre of Excellence, Budapest, Konkoly Thege Mikl\'os \'ut 15-17., H-1121, Hungary} 

\author{P.~L. Palle} 
\affiliation{Instituto de Astrof\'isica de Canarias, 38200 La Laguna, Tenerife, Spain} 
\affiliation{Universidad de La Laguna (ULL), Departamento de Astrof\'isica, 38206 La Laguna, Tenerife, Spain}        

\author{A. P\'al} 
\affiliation{Konkoly Observatory, HUN-REN Research Centre for Astronomy and Earth Sciences, Konkoly Thege Mikl\'os \'ut 15-17, H-1121 Budapest, Hungary}
\affiliation{CSFK, MTA Centre of Excellence, Budapest, Konkoly Thege Mikl\'os \'ut 15-17., H-1121, Hungary}

\author{L. Szigeti} 
\affiliation{ELTE E\"otv\"os Lor\'and University, Gothard Astrophysical Observatory, Szent Imre h. u. 112, 9700, Szombathely, Hungary} 
\affiliation{MTA-ELTE Lend\"ulet "Momentum" Milky Way Research Group, Hungary} 

\author{Sz. M\'esz\'aros} 
\affiliation{ELTE E\"otv\"os Lor\'and University, Gothard Astrophysical Observatory, Szent Imre h. u. 112, 9700, Szombathely, Hungary}
\affiliation{MTA-ELTE Lend\"ulet "Momentum" Milky Way Research Group, Hungary}

\begin{abstract}

Understanding the atmospheric parameters of stars on the top of the RGB is essential to reveal the chemical composition of the Milky Way, as they can be used to probe the farthest parts of our Galaxy. Our goal is to determine the chemical composition of 21 RGB stars with $T_{\mathrm{eff}}<4200$\,K selected from the APOGEE-2 DR17 database using new observations carried out with the spectrograph mounted on the 1-m telescope of the Hungarian Piszk\'estet\H{o} Observatory and the SONG spectrograph (R=77\,000) on the Hertzsprung SONG telescope in the 4500$-$5800\,\r{A} wavelength range. This is the first time the spectrograph (R=18\,000) on the 1-m telescope at Piszk\'estet\H{o} Observatory was used to measure the abundances of stars. We created a new LTE spectral library using MARCS model atmospheres and \texttt{SYNSPEC} by including the line list of 23 molecules to determine atmospheric parameters ($T_{\mathrm{eff}}$, $\log g$, [Fe/H], [$\alpha$/Fe]) and abundances of Si, Ca, Ti, V, Cr, Mn, and Ni with \texttt{FERRE}. The resulting parameters were compared to that of APOGEE. We found a good agreement in general, the average difference is $-$11.2\,K in $T_{\mathrm{eff}}$, 0.11\,dex in $\log g$, 0.10\,dex in [Fe/H], and $-$0.01\,dex in [$\alpha$/Fe]. Our analysis successfully demonstrates the ability of the spectrograph at Piszk\'estet\H{o} Observatory to reliably measure the abundance of bright stars.

\end{abstract}


\section{Introduction}

The general objective of large spectroscopic sky surveys is the mapping of the chemical composition of the Milky Way. Studying the distribution of chemical elements can help us to understand the formation, as well as the evolution of our Galaxy and its substructures. Today's most significant high-resolution spectroscopic sky surveys are the Apache Point Observatory Galactic Evolution Experiment (APOGEE, \citealp{Majewskietal2017}) within the third and fourth phase of Sloan Digital Sky Survey (SDSS, \citealp{Eisensteinetal2011,Blantonetal2017}), the Gaia-ESO Survey (GES, \citealp{Gilmoreetal2012}), and the Galactic Archaeology with HERMES (GALAH, \citealp{DeSilvaetal2015}). The total amount of stars whose spectra are observed and analyzed is more than a million in all programs combined. Upcoming surveys, like the successor to the APOGEE, Milky Way Mapper (MWM, \citealp{Kollmeieretal2017}), 4MOST and WEAVE \citep{deJongetal2019,Daltonetal2016} will observe the spectra of several millions of stars in the Galaxy both in the optical and in the near-infrared H-band by 2030. 

In APOGEE's last data release (DR17, \citealp{Abdurroufetal2022}), the main atmospheric parameters of about 700\,000 stars have been published, as well as the abundances of 20 chemical species (e.g., \citealp{Eilersetal2022,Myersetal2022}). The APOGEE spectrographs \citep{Wilsonetal2019} have a resolution of R$\sim$22\,500 observing in the near infrared region (15\,140$-$16\,900\,\r{A}), which allows to map the Milky Way's central and bulge regions covered by dust \citep{BelokurovKravtsov2022}. The GALAH+ survey took spectra in the optical range between 4713 and 7887\,\r{A}, and observed 580\,000 targets with an R$\sim$28\,000 spectral resolution \citep{Buderetal2021}. GES also operated in the optical wavelength range between 4033 and 9001\,\r{A}. The GIRAFFE and the UVES spectrographs \citep{Pasquinietal2002} used by GES have a spectral resolution between 16\,000 and 47\,000. The GES program observed ca. 115\,000 stars \citep{Randichetal2022}. All of these surveys put a significant effort into observing the top of the RGB to cover the most distant parts of the Milky Way.

\cite{Hegedusetal2023} carried out a detailed comparison of the parameters of APOGEE, GALAH, and GES. They found that the atmospheric parameters of these surveys are in good agreement in general. However, there is a significant difference in the parameters at low temperatures ($T_{\mathrm{eff}}<4500$\,K). These surveys recognize the large uncertainties of parameters on the top of the RGB and usually introduce quality control flags to mark these stars. In fact, the flagging systems of GALAH DR3 \citep{Buderetal2022} and GES DR5 \citep{Gilmoreetal2022} marked most of the $T_{\mathrm{eff}}<4000$\,K stars with a bad flag. This does not necessarily mean these parameters are indeed unreliable but indicates that their accuracy and precision are difficult to validate.

The difference of the atmospheric parameters from the different sky surveys is a general problem for the low-temperature stars \citep{Duetal2021,Passeggeretal2022}. The determination of the parameters is complicated due to the modeling difficulties of the molecular-rich cool stellar atmospheres, which requires the use of MARCS model atmospheres \citep{Gustafssonetal2008} and the inclusion of the latest molecular line lists available.

\begin{figure*}
    \begin{center}
      \includegraphics[width=\textwidth]{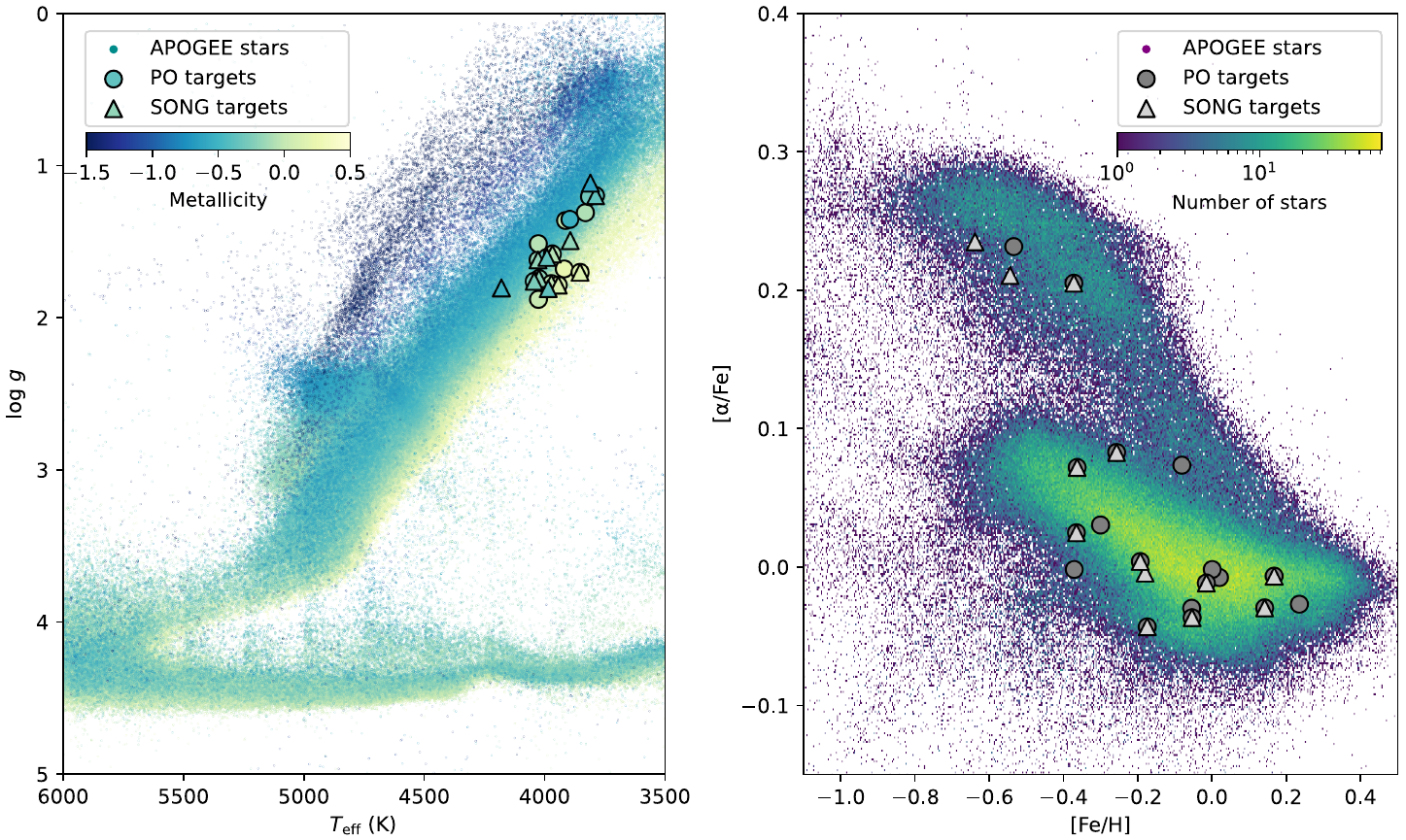}
    \end{center}
    \caption{Left panel: surface gravity as a function of effective temperature of stars. Metallicity is indicated by color shading. Right panel: distribution of the $\alpha$-element abundance as a function of metallicity. The number density of stars is color-shaded. The APOGEE DR17 sample is indicated by dots. The observations of PO and SONG targets are displayed by circles and triangles, respectively. This figure shows APOGEE raw parameters.}
    \label{fig:targets}
\end{figure*}

Apart from the above-mentioned high-resolution spectroscopic sky surveys, studying the chemical composition of cool red giants is a neglected field in astronomy, though deriving independent accurate abundances of these stars in the Milky Way disk is important in order to map the chemical composition of the farthest parts of our Galaxy. Determination of the stellar atmospheric parameters is a complex task in the case of cool stars because of the presence of relatively frequent molecular lines in the atmosphere. These molecular bands cover a wide wavelength range and block a significant fraction of the stellar radiation when $T_{\mathrm{eff}}<4500$\,K and require an extensive and up-to-date molecular line list that is included in the modeling of the spectra of these stars.

Our main goal was to perform an abundance analysis of cool red giants in the optical band independently from all high-resolution spectroscopic sky surveys. As a first step, we selected 21 RGB stars which were observed by the APOGEE survey, obtained high-resolution spectra in the optical wavelengths, and compared our results with the APOGEE parameters. All of our targets were observed with the New Mexico State University (NMSU) 1-m APOGEE telescope spectrograph for bright stars. With our project, we exploited the capabilities of the high-resolution spectrographs mounted on the 1-m telescope at Piszk\'estet\H{o} Observatory (Hungary) and the SONG spectrograph at Teide Observatory (Tenerife). It was the first time for both telescopes that they were used extensively to determine accurate atmospheric parameters and chemical abundances.

The paper is structured as follows. Our new spectral library and applied methods are described in Section\,\ref{sec:methods}. Our results of the comparison of the APOGEE, Piszk\'estet\H{o} and SONG radial velocities, atmospheric parameters and abundances can be found in Sections\,3, 4, and 5. At last, in Section\,\ref{sec:conclusions}, we give a short overview to conclude our study.

\section{Methods}\label{sec:methods}

\subsection{Observations and data reduction}

Our program stars were selected from the APOGEE DR17 \citep{Abdurroufetal2022} using the following criteria for the raw parameters of the APOGEE data sets: $3800\mathrm{\,K}<T_{\mathrm{eff}}<4200$\,K, $1<\mathrm{log}\,g<2$, $-0.1<\mathrm{[Fe/H]}<0.5$, $-0.1<[\alpha\mathrm{/Fe}]<0.3$. Altogether, 21 stars were selected with $\mathrm{G}<8$\,mag using the third data release of Gaia \citep{GaiaCollaboration2023}. Figure\,\ref{fig:targets} shows the surface gravity as a function of effective temperature (Kiel-diagram) and [$\alpha$/M] as a function of metallicity of the APOGEE DR17 stars, and our selected targets. These targets were observed by two different telescopes. The properties of the targets can be seen in Table\,\ref{tab:properties}.

\begin{table*}
    \centering
    \fontsize{8pt}{12pt}\selectfont 
	\caption{Properties and APOGEE raw parameters of the target stars.}
	\begin{tabular}{c c c c r r c c c c c c}
	\hline\hline

    2MASS ID & HD ID & $T_{\mathrm{eff}}$\,(K) & $\log g$ & \multicolumn{1}{c}{[Fe/H]} & \multicolumn{1}{c}{[$\alpha$/Fe]} & G &  G$_{\rm BP}$ & G$_{\rm RP}$ & $t_{\mathrm{exp, PO}}$\,(s) & $t_{\mathrm{exp, SONG}}$\,(s) & Observatory\\
	\hline

    J01094391+3537137   & HD 6860   & 3817 & 1.21 & $-$0.30 & 0.03    & $\cdots$ & $\cdots$ & $\cdots$ & 100      & $\cdots$ & PO\\
    J01301114+0608377   & HD 9138   & 3988 & 1.81 & $-$0.37 & 0.20    & 4.35     & 5.09     & 3.51     & 1800     & 800      & PO, SONG\\
    J04200995+3157113   & HD 27349  & 3975 & 1.78 & $-$0.02 & $-$0.01 & 5.42     & 6.44     & 4.40     & 3600     & 2800     & PO, SONG\\
    J05022869+4104329   & HD 32069  & 3971 & 1.58 & $-$0.06 & $-$0.04 & 3.16      & 4.07     & 2.44     & 600      & 300      & PO, SONG\\
    J05062972+6110109   & HD 32356  & 4047 & 1.76 & $-$0.26 & 0.08    & 5.52     & 6.26     & 4.68     & 3600     & 2400     & PO, SONG\\
    J05544363$-$1146270 & HD 39853  & 3813 & 1.12 & $-$0.64 & 0.23    & 4.99     & 5.88     & 4.05     & $\cdots$ & 1600     & SONG\\
    J06300297+4641079   & HD 45466  & 4026 & 1.74 & $-$0.20 & 0.0     & 5.36     & 6.14     & 4.51     & 3600     & 1200     & PO, SONG\\
    J07363163+4610488   & HD 60437  & 3856 & 1.70 & 0.16    & $-$0.01 & 4.96     & 5.91     & 3.99     & 3600     & 800      & PO, SONG\\
    J07405852+2301067   & HD 61603  & 3946 & 1.79 & 0.14    & $-$0.03 & 5.33     & 6.19     & 4.43     & 3600     & 1200     & PO, SONG\\
    J08555556+1137335   & HD 76351  & 4029 & 1.51 & $-$0.06 & $-$0.03 & 4.99     & 5.72     & 4.16     & 3600     & $\cdots$ & PO\\
    J09413511+3116398   & HD 83787  & 3792 & 1.20 & $-$0.37 & 0.03    & 5.20     & 6.14     & 4.23     & 3600     & 1000     & PO, SONG\\
    J10254427$-$0703358 & HD 90362  & 3834 & 1.31 & $-$0.08 & 0.07    & 4.94     & 5.85     & 3.99     & 3600     & $\cdots$ & PO\\
    J11301888$-$0300128 & HD 99998  & 3916 & 1.36 & $-$0.37 & $-$0.0  & 4.18     & 5.02     & 3.29     & 1800     & $\cdots$ & PO\\
    J13363360+5241148   & HD 118575 & 3922 & 1.68 & 0.23    & $-$0.03 & 6.35     & 7.18     & 5.46     & 7200     & $\cdots$ & PO\\
    J14153968+1910558   & HD 124897 & 4182 & 1.81 & $-$0.55 & 0.21    & $\cdots$ & $\cdots$ & $\cdots$ & $\cdots$ & 5        & SONG\\
    J14490671+5413537   & HD 131005 & 3898 & 1.35 & $-$0.53 & 0.23    & 6.48     & 7.29     & 5.59     & 7200     & $\cdots$ & PO\\
    J15261738+3420095   & HD 137704 & 3993 & 1.60 & $-$0.36 & 0.07    & 4.94     & 5.71     & 4.09     & 3600     & 800      & PO, SONG\\
    J16131544+0501160   & HD 145892 & 4064 & 1.92 & 0.0     & $-$0.01 & 4.93     & 5.72     & 4.06     & 3600     & $\cdots$ & PO\\
    J17214533+5325135   & HD 157681 & 4030 & 1.62 & $-$0.18 & $-$0.04 & 5.15     & 5.94     & 4.28     & 3600     & 2000     & PO, SONG\\
    J19363755+4830583   & HD 185396 & 4029 & 1.88 & 0.02    & $-$0.01 & 6.49     & 7.26     & 5.64     & 7200     & $\cdots$ & PO\\
    J22290798+0907446   & HD 213119 & 3897 & 1.49 & $-$0.18 & 0.0     & 4.96     & 5.85     & 4.04     & $\cdots$ & 1600     & SONG\\
	\hline
		
	\end{tabular}
    \tablecomments{G, G$_{\rm BP}$, and G$_{\rm RP}$ are Gaia DR3 magnitude values \citep{GaiaCollaboration2023}}
	\label{tab:properties}
\end{table*}

One of the applied telescopes was the 1-m primary mirror RCC telescope at Piszk\'estet\H{o} Observatory (PO, altitude: 944\,m). The RCC telescope is equipped with an R$\sim16\,000-20\,000$ nominal instrumental resolution echelle spectrograph, which is sensitive in the optical wavelength range from 4200\,\r{A} to 8500\,\r{A}. For this study, at least 70 signal-to-noise ratio per pixel (SNR) is required to determine the stellar atmospheric parameters with a sufficiently high accuracy. Our program does not include stars with $T_{\mathrm{eff}}<3800$\,K, because they are too faint to be measured with a sufficiently high SNR with the PO telescope. The exposure time was calculated to reach the minimum of the above-mentioned SNR requirement. The applied exposure times can be seen in Table\,\ref{tab:properties} for each observed star. As nearly constant atmospheric conditions are required during the exposures, the targets were observed near the zenith at the lowest airmass as possible to minimize Earth's atmospheric effects on the observed fluxes. We observed 18 RGB stars at PO. 

Observations were also carried out with the 1-m Hertzsprung SONG (Stellar Observations Network Group) Telescope located at Observatorio del Teide (Tenerife, altitude: 2390\,m). The telescope is equipped with an R$\sim30\,000-112\,000$ high-resolution spectrograph (R=77\,000 was applied for our study) operating at 4400$-$6900\,\r{A} wavelengths. From our program stars, 13 stars were observed by the SONG. In this study, spectra of 21 individual stars were analyzed, 10 of these were observed in both observatories.

Flux standard stars are usually O, B, or A type stars, whose continuum can be easily determined and contain a relatively few atomic lines. $\alpha$\,Lyrae, $\theta$\,Crateris, and $\xi^2$\,Ceti standards were observed at PO with 18\,s, 3600\,s, and 900\,s exposure times, respectively. SONG observed $\alpha$\,Lyrae using 20\,s. The scheduling of the observation of standards is based on that the altitude and the observation time of the standard should be relatively close to the observation of the target stars. In this way, the airmass and the weather conditions of the targets and the standard stars were close enough to each other to minimize the effects of the atmosphere on the quality of flux calibration. 

We followed the standard reduction process for both observations. After the standard reduction steps the wavelength of the spectra was calibrated using ThAr lamps. In the next step, we calibrated the flux of the observed stars, order by order, applying \texttt{IRAF} tasks to reduce the curvature of the orders. This procedure was used to connect the adjacent orders to each other and cut out the large flux anomalies from the edges of the orders. The cosmic ray-originated signals represent a significant problem in the case of long exposures by causing unreasonably high flux values in the spectrum. We filtered out the data points with flux values that are at least an order of magnitude higher than the mean flux values in the vicinity of the given data point. The wavelength-calibrated spectra were converted to air wavelength and the barycentric radial velocity correction was also applied by cross-correlating the observed spectra with synthetic ones using \texttt{iSpec}.  

Unfortunately, the sky conditions did not allow an accurate flux calibration, thus in the final step, all flux-calibrated spectra were continuum-normalized. We applied the VWA semi-automatic software package of \cite{Brunttetal2002}, which adjusts the observed continuum to synthetic templates. Finally, for the continuum normalization process the 4500$-$5800\,\r{A} wavelength range was selected to avoid the deepest TiO molecular absorption bands found in the optical band.

\subsection{Fitting of the observed spectra with a synthetic spectral grid}

We determined the atmospheric parameters of our sample of RGB stars by fitting the continuum-normalized spectra with a new theoretical synthetic spectral library. For the fitting, \texttt{FORTRAN90}-based \texttt{FERRE} \citep{AllendePrietoetal2006} optimization code was used, which finds the best fit by comparing the synthetic spectrum with the observed one and carries out $\chi^2$ minimization. The final \texttt{FERRE} atmospheric parameters of the observed stars are associated with the parameters of the best-fitted model. 

A model grid of synthetic spectra is required for the fitting. We employed the BOSZ spectral database \citep{Bohlinetal2017} originally created for the flux calibration of the James Webb Telescope. The BOSZ grid recently underwent significant improvements by using the \texttt{SYNSPEC} \citep{HubenyLanz2011} general spectrum synthesis package and the addition of multiple molecules important in the atmosphere of cool stars. The details are described in M\'esz\'aros et al. (in prep.) and only a short overview of the procedure is given here. The new synthesis is based on the MARCS \citep{Gustafssonetal2008} stellar atmosphere models that were computed for APOGEE DR16 \citep{Jonssonetal2020}. This model atmosphere grid uses the \citet{Grevesseetal2007} solar reference abundance table. The atomic line list was compiled by \citet{HubenyLanz2011}, and no further changes were made.

However, in order to properly model the spectra of cool giants, we included the most up-to-date version of all molecular lists, many using the improvements made by the ExoMol project \citep{Tennysonetal2016,TennysonYurchenko2012}, from Robert Kurucz's website\footnote{Kurucz 1993 models website: \url{http://kurucz.harvard.edu}} in the model spectra. The list consists of the following 23 molecules: AlH, AlO, C$_2$, CaH, CaO, CH, CN, CO, CrH, FeH, H$_2$, H$_2$O, MgH, MgO, NaH, NH, OH, OH$^+$, SiH, SiO, TiH, TiO, and VO. All of the model spectra are calculated with \texttt{SYNSPEC} with the assumption of local thermodynamic equilibrium (LTE) in a spherical atmosphere.

Our synthetic spectra were generated uniformly with R=300\,000 resolution. However, the spectral resolution of the synthetic spectra has to be the same as the resolution of the observed spectra. First, we determined the exact resolution of the PO spectra based on ThAr lamp reference measurements. The resolution of the observed spectra is found to be R=18\,000 on average from the full width of half maximum of selected spectral lines in the ThAr spectrum, thus all synthetic spectra were convolved to this resolution for the PO observations. We also created an R=77\,000 grid for the SONG spectra.

\begin{table}
	\centering
	\caption{Grid properties of the two \texttt{SYNSPEC} synthetic spectral grids}
	\begin{tabular}{c c c c}
	\hline\hline

    \multirow{2}{*}{} & \multicolumn{3}{c}{Grid 1} \\
	\cline{2-4}

	& minimum & maximum & step size \\\hline
    $T_{\mathrm{eff}}$\,(K)               & 3000  & 4000 & 100  \\
    $\log g$                              & $-$0.5  & 2.5  & 0.5  \\
	$\rm [$Fe/H$\rm ]$                    & $-$2.25 & 0.5  & 0.25 \\
	${\rm [}\alpha$/Fe$\rm ]$             & $-$0.25 & 0.5  & 0.25 \\
	$v_{\mathrm{mic}}$\,(kms$^{-1}$)    & 0     & 4    & 2    \\\hline

    \multirow{2}{*}{} & \multicolumn{3}{c}{Grid 2} \\
	\cline{2-4}

	& minimum & maximum & step size \\\hline
    $T_{\mathrm{eff}}$\,(K)               & 3750  & 4500 & 250  \\
	$\log g$                              & $-$0.5  & 2.5  & 0.5  \\
	$\rm [$Fe/H$\rm ]$                    & $-$2.25 & 0.5  & 0.25 \\
	${\rm [}\alpha$/Fe$\rm ]$             & $-$0.25 & 0.5  & 0.25 \\
	$v_{\mathrm{mic}}$\,(kms$^{-1}$)    & 0     & 4    & 2    \\\hline

	\end{tabular}
	\label{tab:grid}
\end{table}

Table\,\ref{tab:grid} shows the parameter coverage of our synthetic spectra calculation. To provide a finer coverage of $T_{\mathrm{eff}}$ parameter, we created two \texttt{SYNSPEC} sub-grids: 1) 3000$-$4000\,K range with 100\,K grid step, and 2) for 3750$-$4500\,K with 250\,K grid step. The microturbulent velocity was varied between 0, 2, and 4\,kms$^{-1}$.

A two-step process was used to determine atmospheric parameters and abundances after the synthetic spectral grids were interpolated to the wavelength scale of the observed spectra. In the first step, the 4500$-$5800\,\r{A} wavelength range was fitted with the above defined grids and the following main atmospheric parameters were derived: $T_{\mathrm{eff}}$, $\log g$, [Fe/H], [$\alpha$/Fe]. To account for systematic errors in the continuum normalization, we multiplied the observed spectra with numbers between 0.9 and 1.1 using 0.01 steps by steps. All of the multiplied versions of the spectra have been fitted with the new \texttt{SYNSPEC} grid, and the best fit with the lowest reduced $\chi^2$ was selected. We calculated the standard deviation of these multiple fittings to estimate our uncertainties.

In the next step, four of the globally fitted parameters were fixed and only the $\alpha$-element abundances were derived by fitting the individual atomic lines of the following elements with \texttt{FERRE}: Si, Ca, Ti, V, Cr, Mn, and Ni. We selected the best-fitted non-blended atomic absorption lines in the 4500$-$5800\,\r{A} wavelength range with an equivalent width larger than 20\,m\r{A}. The list of selected lines can be seen in Table\,\ref{tab:linelist}. The same line sample was used for both PO and SONG observations.

\begin{table*}
    \centering
    \fontsize{9pt}{12pt}\selectfont
	\caption{Lines used for the element abundance analysis.}
	\begin{tabular}{c c c r c c c}
	\hline\hline

	Element & Ion & Wavelength\,(\r{A}) & log(gf) & PO line mask\,(\r{A}) & SONG line mask\,(\r{A}) \\
	\hline

    Si & 1 & 5665.555 & $-$1.754 & 5665.205 $-$ 5665.955 & 5665.405 $-$ 5665.805 \\
    Ca & 1 & 5349.465 & $-$0.428 & 5348.965 $-$ 5350.065 & 5349.165 $-$ 5349.615 \\
    Ti & 1 & 4533.239 &    0.532 & 4532.839 $-$ 4533.589 & 4532.989 $-$ 4533.489 \\
    Ti & 1 & 4999.503 &    0.306 & 4998.953 $-$ 4999.903 & 4999.203 $-$ 4999.853 \\
    Ti & 1 & 5039.957 & $-$1.206 & 5039.657 $-$ 5040.357 & 5039.657 $-$ 5040.207 \\
    V  & 1 & 4594.080 & $-$1.365 & 4593.680 $-$ 4594.380 & 4593.780 $-$ 4594.280 \\
    Cr & 1 & 4652.157 & $-$1.035 & 4651.807 $-$ 4652.507 & 4651.907 $-$ 4652.407 \\
    Cr & 1 & 4718.420 &    0.240 & 4718.070 $-$ 4718.670 & 4718.220 $-$ 4718.620 \\
    Cr & 1 & 5348.315 & $-$1.294 & 5347.865 $-$ 5348.765 & 5348.115 $-$ 5348.615 \\
    Mn & 1 & 5420.425 & $-$2.029 & 5419.975 $-$ 5420.725 & 5420.125 $-$ 5420.625 \\
    Ni & 1 & 5694.983 & $-$0.467 & 5694.433 $-$ 5695.333 & 5694.833 $-$ 5695.133 \\
	\hline

	\end{tabular}
    \tablecomments{Wavelengths and log(gf) values come from \citet{Hubenyetal2021}}
	\label{tab:linelist}
\end{table*}

We selected sections of the spectra that are sensitive to the abundance derived from only a given absorption line of an element. The exact wavelength range of the sections can be found in Table\,\ref{tab:linelist}, where the average widths of the sections are 0.8 and 0.48\,\r{A} for the PO and the SONG spectra, respectively. We separately fitted the different sections with \texttt{FERRE}. In this phase, only the [Fe/H] parameter was left free, while $T_{\mathrm{eff}}$, $\log g$, [$\alpha$/Fe], and $v_{\mathrm{micro}}$ were fixed on that value which was derived from the global fitting. We applied a continuum displacement in each abundance window just as the continuum-normalized observed spectra were multiplied by numbers between 0.9 and 1.1, then all displacement versions were fitted with a 0.01 step size, and the least $\chi^2$ fitting was selected for the analysis. We determined the individual abundance from the difference between the global metallicity determined in the first step, and the new abundance derived in the second step for a given atomic line. If an element had more lines, individual abundances were averaged together for all lines of the given element to get the final abundance (see the results in Section\,\ref{sec:abundance}).

\section{Radial velocities}

\begin{figure}
    \begin{center}
      \includegraphics[width=0.9\columnwidth]{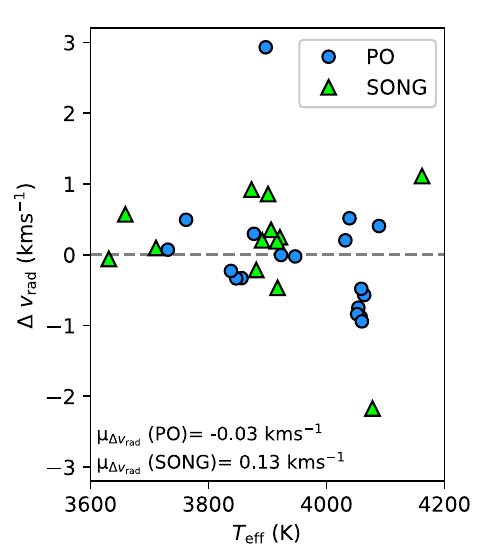}
    \end{center}
    \caption{Radial velocity differences of APOGEE values and our targets as a function of effective temperature. PO and SONG differences are denoted by blue circles and green triangles, respectively.}
    \label{fig:radvel}
\end{figure}

We determined the radial velocities of all PO and SONG targets by applying the cross-correlation method using a well-matching theoretical template spectrum from our \texttt{SYNSPEC} spectral library. We made barycentric corrections to every radial velocity value. The calculated $v_{\mathrm{rad}}$ values are displayed in Table\,\ref{tab:params}. To check the accuracy of these radial velocities, their values are compared with APOGEE ones, which have also been given in the barycentric frame of the Solar System. The $v_{\mathrm{rad}}$ differences ($\Delta v_{\mathrm{rad}}$) between APOGEE values and our results are shown in Figure\,\ref{fig:radvel} as a function of effective temperature.  

The precision of the APOGEE $v_{\mathrm{rad}}$ is close to 0.05\,kms$^{-1}$, which can be better for brighter, cooler, and more metal-rich stars \citep{Nideveretal2015}. Our $\Delta v_{\mathrm{rad}}$ values are between $-$1 and 1\,kms$^{-1}$ for all PO and SONG targets, with one outlier for both telescopes: $\Delta v_{\mathrm{rad}} \sim 3$\,kms$^{-1}$ for one PO target, and $\Delta v_{\mathrm{rad}} \sim -2$\,kms$^{-1}$ for one SONG target. The average differences are $-$0.03 and 0.13\,kms$^{-1}$ including all PO and SONG observations, respectively. The standard deviation of $\Delta v_{\mathrm{rad}}$ values are 0.85 and 0.79\,kms$^{-1}$ for PO and SONG, respectively, based on the $v_{\mathrm{rad}}$ values of Table\,\ref{tab:params}. The average error of PO and SONG $v_{\mathrm{rad}}$ values are 0.27 and 0.06\,kms$^{-1}$, respectively. Our average $v_{\mathrm{rad}}$ errors are larger than the average differences, thus, we can reproduce the APOGEE $v_{\mathrm{rad}}$ values with our observations within the calculated uncertainties.

\section{Atmospheric parameters}\label{sec:properties}

\begin{figure*}
    \begin{center}
      \includegraphics[width=\textwidth]{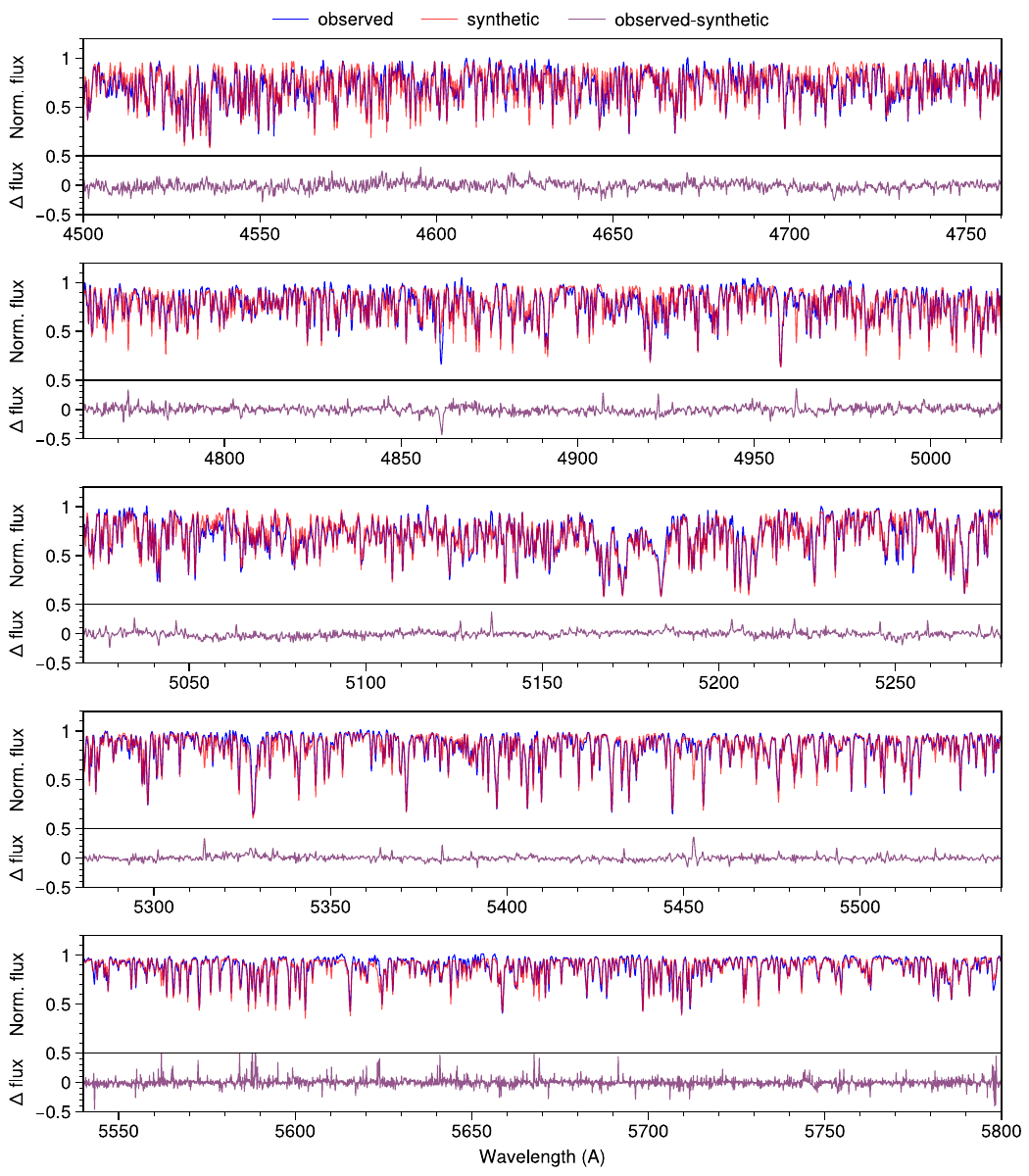}
    \end{center}
    \caption{Continuum normalized high-resolution spectrum of HD\,32356 in the 4500$-$5800\,\r{A} wavelength range observed with the PO spectrograph (blue line) fitted with a synthetic spectrum (red line) using the \texttt{FERRE} code. The observed$-$synthetic flux difference is indicated with the purple line.}
    \label{fig:fittings}
\end{figure*}

In this section, the accuracy and precision of atmospheric parameters derived as part of the global fittings are discussed for stellar spectra of our targets. Figure\,\ref{fig:fittings} is meant to present the general quality of the global spectrum fittings with synthetic spectra for a selected cool giant (HD\,32356) in the 4500$-$5800\,\r{A} wavelength range. Figure\,\ref{fig:fittings} also shows the difference between the observed and synthetic spectra. The small oscillation in the difference around zero shows a very good quality of the fitting. 

\begin{figure*}
    \begin{center}
      \includegraphics[width=0.9\textwidth]{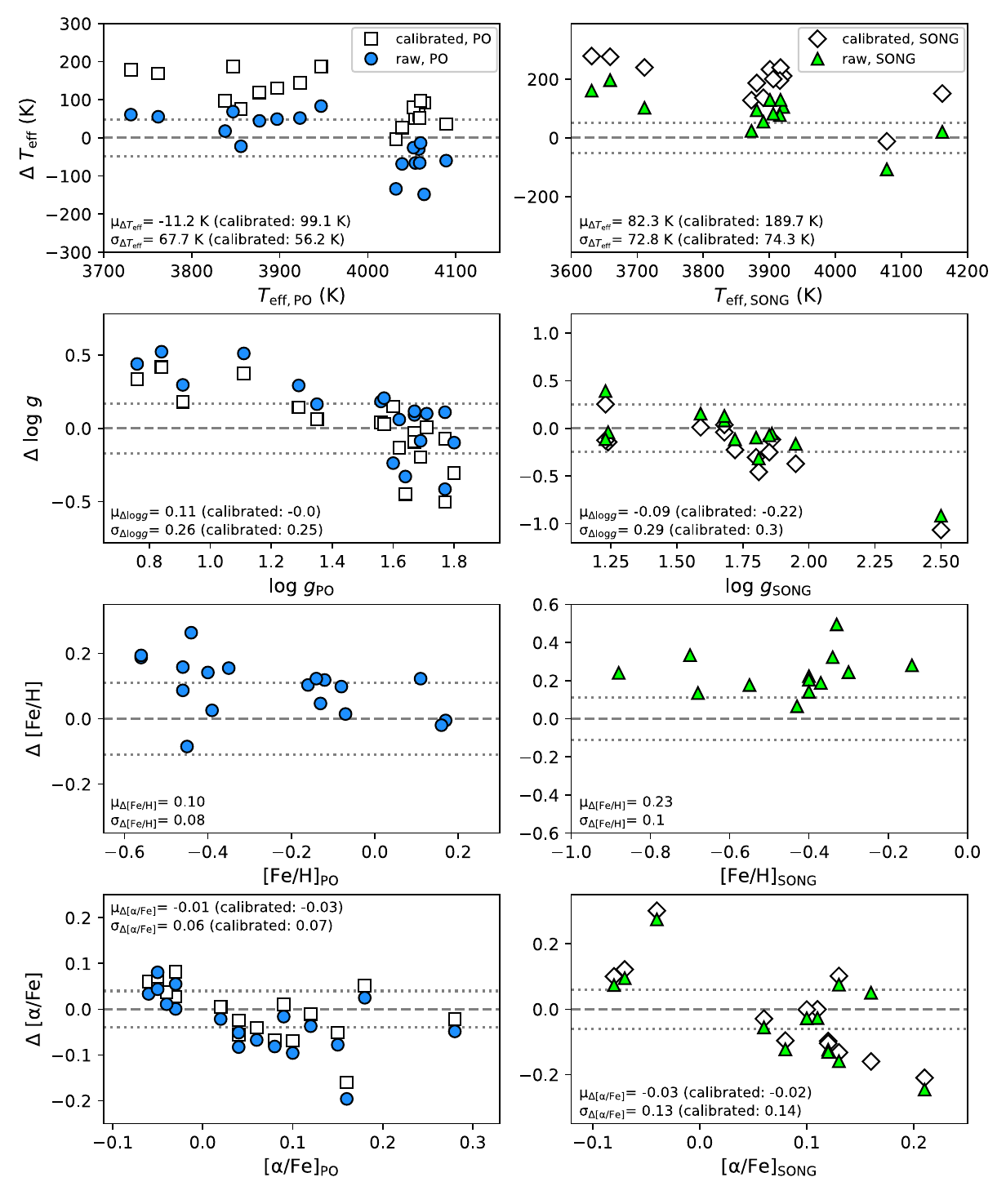}
    \end{center}
    \caption{The differences between the APOGEE and our fitted parameters as a function of stellar atmospheric parameters of this study. Left and right panels display PO and SONG parameters, respectively, 1$^{\mathrm{st}}$ row: effective temperature; 2$^{\mathrm{nd}}$ row: surface gravity; 3$^{\mathrm{rd}}$ row: metallicity; 4$^{\mathrm{th}}$ row: $\alpha$-element abundance. Blue filled circles and empty squares are related to the raw and the calibrated APOGEE$-$PO values, respectively. Green filled triangles and empty rhombi denote the raw and the calibrated APOGEE$-$SONG values, respectively. The dotted grey lines belong to the upper and lower average error limits. The average and the standard deviation of the differences are denoted by $\mu$ and $\sigma$ symbols, respectively.}
    \label{fig:synple_panels}
\end{figure*}

The total list of the best-fitted parameters and their uncertainties can be found in Table\,\ref{tab:params} for 18 PO and 13 SONG observations as well. Our main goal was the validation of the APOGEE parameters using the PO and SONG spectrographs. The agreement can be considered good if the parameters of the two sources agree within the reported uncertainties. APOGEE has published both raw and calibrated $T_{\mathrm{eff}}$ and $\log g$ values, and because our derived parameters are purely spectroscopic, we chose the raw (spectroscopic) APOGEE parameters for the primary comparison. However, comparisons with the calibrated APOGEE parameters were also made for completeness. Figure\,\ref{fig:synple_panels} shows the differences between the APOGEE and PO, as well as SONG parameters as a function of our atmospheric parameters. The estimated uncertainties of $T_{\mathrm{eff}}$, $\log g$, [Fe/H], and [$\alpha$/Fe] are derived from the mean error of the fitted parameters added in quadrature to the APOGEE errors. The common uncertainties are denoted by dotted horizontal lines in Figure\,\ref{fig:synple_panels} to indicate the region in which the differences can be generally considered good.

\subsection{Effective temperature, $T_{\mathrm{eff}}$}

The effective temperature of the 18 PO targets is compared to the APOGEE values in the top left panel of Figure\,\ref{fig:synple_panels} for both the raw and calibrated APOGEE parameters. The APOGEE raw $T_{\mathrm{eff}}$ is derived directly from the H-band spectra, and its accuracy is checked against the temperature determined with the infrared-flux method (IRFM, \citealp{Casagrande2008}) scale in DR17. The small offsets were calibrated to the IRFM scale using the calibration relations of \citet{GonzalezHernandezBonifacio2009}, resulting in the second temperature scale of APOGEE.

We look at the differences between the raw APOGEE and PO temperatures ($\Delta T_{\mathrm{eff}}$) first. The average $\Delta T_{\mathrm{eff}}$ is $-$11.2\,K for the raw APOGEE values with a 67.7\,K of standard deviation. The average error of PO $T_{\mathrm{eff}}$ is 47.5\,K. These differences are considered to be within both our and APOGEE's uncertainties. We can observe a duality in the $T_{\mathrm{eff}}$ differences: for $T_{\mathrm{eff}}<4000$\,K the APOGEE temperatures are slightly lower than the ones obtained from PO, and for $T_{\mathrm{eff}}>4000$\,K the opposite can be observed, however, these are both within our uncertainties. Based on our results, we can conclude that our fitted effective temperatures show a very good agreement with the spectroscopic raw APOGEE $T_{\mathrm{eff}}$ values.

In the case of calibrated APOGEE parameters, $\Delta T_{\mathrm{eff}}$ values are outside the tolerance range for $T_{\mathrm{eff}}<4000$\,K stars, however, they are inside the tolerance range in the case of stars above 4000\,K. The average $\Delta T_{\mathrm{eff}}$ is 99.1\,K for the calibrated values, and the standard deviation is 56.2\,K. Thus, the $\Delta T_{\mathrm{eff}}$ values are more significant for the calibrated than for the raw APOGEE values. The consistently higher differences for the calibrated values can be explained by the fact that the \citet{GonzalezHernandezBonifacio2009} photometric $T_{\mathrm{eff}}$ values are valid between 4500 and 7000\,K, that is outside of our temperature range. The APOGEE calibrated values below 4500\,K are calculated with extrapolation from the end of the calibrated range, which appears to result in erroneous APOGEE calibrated temperatures.

We also compared the APOGEE and the SONG $T_{\mathrm{eff}}$ values for 13 targets. The APOGEE$-$SONG $T_{\mathrm{eff}}$ differences are shown in the top right panel of Figure\,\ref{fig:synple_panels} for raw and calibrated values. The average $\Delta T_{\mathrm{eff}}$ is 82.3\,K for the raw APOGEE values, and the standard deviation of $\Delta T_{\mathrm{eff}}$ is 72.8\,K. The average SONG $T_{\mathrm{eff}}$ uncertainty is found to be 51.6\,K. Most of the SONG $\Delta T_{\mathrm{eff}}$ values are within the APOGEE error limits for the raw parameters. In the case of calibrated APOGEE $T_{\mathrm{eff}}$ values, the average $\Delta T_{\mathrm{eff}}$ is 189.7\,K with a 74.3\,K of standard deviation. As one can see, all of the APOGEE$-$SONG differences are above the upper APOGEE uncertainty limit (excepting only one $\Delta T_{\mathrm{eff}}$ value). However, the consistency with the raw spectroscopic APOGEE parameters seems to be quite good, the APOGEE$-$SONG average differences are higher than the APOGEE$-$PO values. It can be explained by larger inaccuracies during the SONG data reduction: the higher resolution spectra of the SONG spectrograph require greater caution during continuum normalization. However, the fact that the general SNR level of the SONG spectra is slightly lower than that of PO, can play a role in the larger uncertainties in the case of SONG results. 

Comparing the fitted parameters of the 10 shared targets, it can be seen that $T_{\mathrm{eff}}$ values of SONG are lower than PO values (see Table\,\ref{tab:params}). The difference between the two observations is $-$89.9\,K on the average. The standard deviation of SONG$-$PO $\Delta T_{\mathrm{eff}}$ is 68.3\,K. Since the average error of $T_{\mathrm{eff}}$ is 47.5 and 51.6\,K for PO and SONG, respectively, most of the $T_{\mathrm{eff}}$ differences are within the PO and SONG uncertainties. We can conclude that the PO $T_{\mathrm{eff}}$ values are slightly larger than the SONG parameters, but the difference is not significant considering the range of our estimated errors.

\begin{table*}
    \centering
    \fontsize{9pt}{12pt}\selectfont
	\caption{Atmospheric parameters of the stars observed with PO RCC and SONG telescopes.}
    \begin{tabular}{c l l r r l r}
	\hline\hline

    \multirow{2}{*}{{2MASS ID}} & \multicolumn{6}{c}{PO} \\
	\cline{2-7}
	
    & \multicolumn{1}{c}{$T_{\mathrm{eff}}$\,(K)} & \multicolumn{1}{c}{$\log g$} & \multicolumn{1}{c}{[Fe/H]} & \multicolumn{1}{c}{[$\alpha$/Fe]} & \multicolumn{1}{c}{$v_{\mathrm{mic}}$\,(kms$^{-1}$)} & \multicolumn{1}{c}{$v_{\mathrm{rad}}$\,(kms$^{-1}$)}\\
	\hline
    J01094391+3537137 & 3762 $\pm$ 40.4 & 0.91 $\pm$ 0.10 & $-$0.46 $\pm$ 0.03 & $-$0.05 $\pm$ 0.04 & 1.10 $\pm$ 0.03 & 0.16   $\pm$ 0.26\\
    J01301114+0608377 & 4054 $\pm$ 45.3 & 1.71 $\pm$ 0.15 & $-$0.46 $\pm$ 0.11 & 0.18  $\pm$ 0.04 & 0.84 $\pm$ 0.07 & 35.35  $\pm$ 0.24\\
    J04200995+3157113 & 3923 $\pm$ 59.9 & 1.57 $\pm$ 0.21 & $-$0.14 $\pm$ 0.13 & 0.04  $\pm$ 0.05 & 1.04 $\pm$ 0.06 & $-$18.32 $\pm$ 0.26\\
    J05022869+4104329 & 4039 $\pm$ 59.4 & 1.29 $\pm$ 0.27 & $-$0.07 $\pm$ 0.12 & 0.16  $\pm$ 0.03 & 0.74 $\pm$ 0.07 & $-$3.93  $\pm$ 0.29\\
    J05062972+6110109 & 4060 $\pm$ 43.1 & 1.67 $\pm$ 0.15 & $-$0.40 $\pm$ 0.11 & 0.12  $\pm$ 0.02 & 0.83 $\pm$ 0.07 & $-$44.62 $\pm$ 0.25\\
    J06300297+4641079 & 4052 $\pm$ 29.0 & 1.56 $\pm$ 0.12 & $-$0.35 $\pm$ 0.13 & 0.10  $\pm$ 0.02 & 0.96 $\pm$ 0.09 & $-$47.18 $\pm$ 0.25\\
    J07363163+4610488 & 3838 $\pm$ 88.3 & 1.80 $\pm$ 0.24 & 0.17  $\pm$ 0.10 & $-$0.05 $\pm$ 0.07 & 0.92 $\pm$ 0.03 & 26.33  $\pm$ 0.34\\
    J07405852+2301067 & 3897 $\pm$ 74.5 & 1.67 $\pm$ 0.21 & 0.16  $\pm$ 0.09 & $-$0.04 $\pm$ 0.05 & 0.94 $\pm$ 0.04 & 36.69  $\pm$ 0.28\\
    J08555556+1137335 & 4089 $\pm$ 18.9 & 1.35 $\pm$ 0.12 & $-$0.16 $\pm$ 0.11 & $-$0.03 $\pm$ 0.04 & 1.00 $\pm$ 0.09 & 25.42  $\pm$ 0.25\\
    J09413511+3116398 & 3731 $\pm$ 36.2 & 0.76 $\pm$ 0.17 & $-$0.56 $\pm$ 0.10 & $-$0.03 $\pm$ 0.04 & 1.01 $\pm$ 0.07 & $-$16.13 $\pm$ 0.28\\
    J10254427$-$0703358       & 3856 $\pm$ 64.1 & 1.64 $\pm$ 0.22 & $-$0.13 $\pm$ 0.10 & 0.09  $\pm$ 0.07 & 0.73 $\pm$ 0.05 & 36.13  $\pm$ 0.36\\
    J11301888$-$0300128       & 3847 $\pm$ 78.0 & 0.84 $\pm$ 0.35 & $-$0.56 $\pm$ 0.20 & 0.02  $\pm$ 0.07 & 1.07 $\pm$ 0.14 & 18.79  $\pm$ 0.24\\
    J13363360+5241148 & 3877 $\pm$ 61.4 & 1.62 $\pm$ 0.24 & 0.11  $\pm$ 0.13 & $-$0.06 $\pm$ 0.04 & 0.86 $\pm$ 0.06 & $-$17.33 $\pm$ 0.32\\
    J14490671+5413537 & 4032 $\pm$ 28.7 & 1.77 $\pm$ 0.08 & $-$0.45 $\pm$ 0.12 & 0.28  $\pm$ 0.07 & 0.90 $\pm$ 0.08 & $-$50.51 $\pm$ 0.23\\
    J15261738+3420095 & 4059 $\pm$ 57.6 & 1.69 $\pm$ 0.20 & $-$0.39 $\pm$ 0.11 & 0.15  $\pm$ 0.03 & 0.84 $\pm$ 0.08 & $-$48.32 $\pm$ 0.24\\
    J16131544+0501160 & 4064 $\pm$ 12.2 & 1.60 $\pm$ 0.06 & $-$0.12 $\pm$ 0.09 & 0.08  $\pm$ 0.02 & 0.84 $\pm$ 0.06 & $-$0.53  $\pm$ 0.26\\
    J17214533+5325135 & 3947 $\pm$ 45.2 & 1.11 $\pm$ 0.16 & $-$0.44 $\pm$ 0.12 & 0.04  $\pm$ 0.03 & 1.10 $\pm$ 0.08 & $-$7.10  $\pm$ 0.24\\
    J19363755+4830583 & 4058 $\pm$ 13.6 & 1.77 $\pm$ 0.07 & $-$0.08 $\pm$ 0.09 & 0.06  $\pm$ 0.02 & 0.81 $\pm$ 0.06 & $-$6.76  $\pm$ 0.27\\
	\hline
	
    \multirow{2}{*}{{2MASS ID}} & \multicolumn{6}{c}{SONG} \\
	\cline{2-7}
	
    & \multicolumn{1}{c}{$T_{\mathrm{eff}}$\,(K)} & \multicolumn{1}{c}{$\log g$} & \multicolumn{1}{c}{[Fe/H]} & \multicolumn{1}{c}{[$\alpha$/Fe]} & \multicolumn{1}{c}{$v_{\mathrm{mic}}$\,(kms$^{-1}$)} & \multicolumn{1}{c}{$v_{\mathrm{rad}}$\,(kms$^{-1}$)}\\
	\hline
    J01301114+0608377 & 3906 $\pm$ 97.1  & 1.68 $\pm$ 0.29 & $-$0.55 $\pm$ 0.1  & 0.13  $\pm$ 0.1  & 1.24 $\pm$ 0.05 & 34.25  $\pm$ 0.05\\
    J04200995+3157113 & 3881 $\pm$ 6.8   & 1.85 $\pm$ 0.03 & $-$0.34 $\pm$ 0.03 & 0.12  $\pm$ 0.02 & 2.49 $\pm$ 0.02 & $-$18.11 $\pm$ 0.06\\
    J05022869+4104329 & 4078 $\pm$ 93.5  & 2.5  $\pm$ 0.33 & $-$0.3  $\pm$ 0.13 & 0.21  $\pm$ 0.1  & 2.87 $\pm$ 0.01 & $-$1.24  $\pm$ 0.09\\
    J05062972+6110109 & 3917 $\pm$ 102.5 & 1.68 $\pm$ 0.33 & $-$0.4  $\pm$ 0.18 & 0.11  $\pm$ 0.1  & 1.15 $\pm$ 0.41 & $-$45.09 $\pm$ 0.06\\
    J05544363$-$1146270       & 3711 $\pm$ 48.3  & 1.23 $\pm$ 0.32 & $-$0.88 $\pm$ 0.11 & $-$0.04 $\pm$ 0.08 & 2.59 $\pm$ 0.48 & 81.41  $\pm$ 0.06\\
    J06300297+4641079 & 3921 $\pm$ 103.9 & 1.59 $\pm$ 0.4  & $-$0.4  $\pm$ 0.2  & 0.06  $\pm$ 0.06 & 1.27 $\pm$ 0.41 & $-$48.27 $\pm$ 0.06\\
    J07363163+4610488 & 3659 $\pm$ 63.8  & 1.8  $\pm$ 0.3  & $-$0.33 $\pm$ 0.11 & $-$0.08 $\pm$ 0.01 & 2.5  $\pm$ 0.05 & 25.53  $\pm$ 0.09\\
    J07405852+2301067 & 3891 $\pm$ 12.3  & 1.95 $\pm$ 0.17 & $-$0.14 $\pm$ 0.08 & 0.13  $\pm$ 0.05 & 2.65 $\pm$ 0.02 & 39.42  $\pm$ 0.07\\
    J09413511+3116398 & 3631 $\pm$ 3.8   & 1.24 $\pm$ 0.34 & $-$0.7  $\pm$ 0.09 & $-$0.07 $\pm$ 0.07 & 2.49 $\pm$ 0.05 & $-$16.0 $\pm$ 0.07\\
    J14153968+1910558 & 4162 $\pm$ 20.3  & 1.86 $\pm$ 0.08 & $-$0.68 $\pm$ 0.1  & 0.16  $\pm$ 0.01 & 1.36 $\pm$ 0.47 & $-$5.74  $\pm$ 0.06\\
    J15261738+3420095 & 3916 $\pm$ 91.8  & 1.72 $\pm$ 0.33 & $-$0.43 $\pm$ 0.14 & 0.1   $\pm$ 0.1  & 1.2  $\pm$ 0.4  & $-$48.99 $\pm$ 0.05\\
    J17214533+5325135 & 3901 $\pm$ 9.5   & 1.23 $\pm$ 0.14 & $-$0.4  $\pm$ 0.09 & 0.08  $\pm$ 0.07 & 1.31 $\pm$ 0.39 & $-$7.98  $\pm$ 0.05\\
    J22290798+0907446 & 3873 $\pm$ 17.8  & 1.81 $\pm$ 0.21 & $-$0.37 $\pm$ 0.09 & 0.12  $\pm$ 0.07 & 2.46 $\pm$ 0.02 & $-$30.89 $\pm$ 0.06\\
	\hline
	
	\end{tabular}
	\label{tab:params}
\end{table*}

\subsection{Surface gravity, $\log g$}

The determination of $\log g$ is often complicated, therefore it is less precise than other atmospheric parameters, and usually its estimated uncertainty is $0.1-0.3$\,dex. Our fitted $\log g$ values and errors of PO and SONG spectra are displayed in Table\,\ref{tab:params}. The estimated average errors of our PO and SONG spectroscopic $\log g$ values are 0.17 and 0.25\,dex, respectively. The average difference of the SONG$-$PO $\log g$ is 0.24\,dex for the 10 shared targets. The standard deviation of SONG$-$PO $\log g$ is 0.36\,dex. The SONG$-$PO $\log g$ values are close to the error limits of the two telescopes. 

The 2$^{\mathrm{nd}}$ row left panel of Figure\,\ref{fig:synple_panels} shows the APOGEE$-$PO $\log g$ differences as a function of PO $\log g$ separately for raw and calibrated APOGEE parameters. The calibration of APOGEE DR17 $\log g$ is based on the comparison of $\log g$ values from asteroseismology \citep{Serenellietal2017} and isochrones \citep{Bergeretal2020}. In order to calibrate $\log g$, a neural network was applied to eliminate the discontinuities between the different groups of stars, which was used in APOGEE DR16 calibration too. The asteroseismic surface gravity values provide an excellent comparison as their uncertainties are an order of magnitude smaller than spectroscopic $\log g$ values \citep{Pinsonneaultetal2018}.

One can also see a duality in $\Delta$$\log g$  differences similarly to $\Delta T_{\mathrm{eff}}$: the two clumps are at log\,$g<0.15$\,dex as well as at log\,$g>0.15$\,dex. The average raw $\Delta$$\log g$ is 0.11\,dex, and the standard deviation of the raw difference is 0.26\,dex. Our surface gravity values agree better with the calibrated $\Delta$$\log g$ values, as the average calibrated $\Delta$$\log g$ is zero with a 0.25\,dex of standard deviation. 

We can conclude that most of our fitted $\log g$ values match both the raw and the calibrated APOGEE parameters between 1.5 and 1.8\,dex (see top right panel of Fig\,\ref{fig:synple_panels}). Moreover, the differences in the calibrated values are lower than in the raw values. We emphasize in the context of calibrated APOGEE values that the neural network was trained on ASPCAP stars, where the lowest $\log g$ value is $\sim2.2$\,dex. Thus, the calibration may become uncertain below 2.2\,dex. 

The comparison of the 13 SONG and APOGEE $\log g$ parameters are shown in the 2$^{\mathrm{nd}}$ row right panel of Figure\,\ref{fig:synple_panels} for raw and calibrated values. We can also observe the two clumps in the APOGEE$-$SONG differences below and above 0.15\,dex in $\log g$. The $\Delta$$\log g$ is $-$0.09\,dex on the average for the raw APOGEE values, and the standard deviation of $\Delta$$\log g$ values is 0.29\,dex. The differences for the calibrated APOGEE $\log g$ values are found to be slightly lower than for the raw values, the average $\Delta$$\log g$ is $-$0.22\,dex, and the standard deviation is 0.3\,dex. Both for raw and calibrated APOGEE values, the matching of SONG and APOGEE $\log g$ values is very well considering the estimated uncertainties.

In the full $\log g$ region, one can also observe a linear trend in the $\Delta$$\log g$ values, similarly to $\Delta T_{\mathrm{eff}}$ values: $\Delta$$\log g$ is decreasing with the $\log g$ from PO and SONG spectra, this trend may be caused by systematics in our fitting procedure. However, both PO and SONG standard deviations of $\Delta$$\log g$ are close to the APOGEE error range.

\subsection{Abundance parameters: [Fe/H] and [$\alpha$/Fe]}

The [Fe/H] and [$\alpha$/Fe] values from the PO and the SONG spectra are shown in Table\,\ref{tab:params}. The average difference between PO and SONG [Fe/H] values is $-$0.15\,dex. The average [Fe/H] error is 0.11\,dex for both PO and SONG observations. The standard deviation of SONG$-$PO [Fe/H] is 0.15\,dex. Most of the [Fe/H] differences are what is expected from the error range of the PO telescope. The [$\alpha$/Fe] abundances show excellent agreement between the two telescopes, the average difference is very close to zero. The average [$\alpha$/Fe] error is 0.04 and 0.06\,dex for PO and SONG, respectively. The standard deviation of [$\alpha$/Fe] differences is found to be 0.07\,dex. Based on the comparison of fitted [Fe/H] and [$\alpha$/Fe] parameters, we conclude that the PO observations can match the SONG parameters within the expected uncertainties.

The 3$^{\mathrm{rd}}$ row left panel of Figure\,\ref{fig:synple_panels} presents the APOGEE$-$PO [Fe/H] differences as a function [Fe/H] of PO observations. In the case of [Fe/H], APOGEE did not perform any calibrations. We found an average $\Delta$[Fe/H] systematic offset of 0.1\,dex with 0.08\,dex of standard deviation. This systematic error may be the result of minor continuum normalization inaccuracies of the optical spectra because of the molecular bands appearing with decreasing temperature. While a slight systematic offset appears in our analysis, it is fairly small and may disappear at higher temperatures where the continuum normalization can be carried out more precisely.

The 3$^{\mathrm{rd}}$ right panel of Figure\,\ref{fig:synple_panels} shows the comparison of the SONG and the spectroscopic APOGEE [Fe/H] values. The average [Fe/H] difference is 0.23\,dex, and the standard deviation is 0.1\,dex. One can see that except for one $\Delta$[Fe/H] value, all differences are outside both APOGEE and SONG estimated uncertainties. The SONG [Fe/H] values underestimate APOGEE parameters even more than the PO values. Metallicity is one of the atmospheric parameters that is very sensitive to the accuracy of the observation and the continuum normalization. Thus, the [Fe/H] systematic uncertainty is more apparent at such a high spectral resolution that SONG has.

The bottom left panel of Figure\,\ref{fig:synple_panels} displays the comparison of the PO and the raw, as well as the calibrated [$\alpha$/Fe] APOGEE values. In the case of giants, the calibrated [$\alpha$/Fe] is based on a small zero-point shift of 0.03\,dex in order to force the [$\alpha$/Fe] to be zero for the stars with a solar-metallicity in the solar neighborhood \citep{Jonssonetal2020}. The average raw $\Delta$[$\alpha$/Fe] is $-$0.01\,dex, the standard deviation of raw $\Delta$[$\alpha$/Fe] is 0.06\,dex. For the calibrated APOGEE parameters, the average $\Delta$[$\alpha$/Fe] is $-$0.03, and the standard deviation is 0.07\,dex. The raw and the calibrated [$\alpha$/Fe] differences are generally very close to each other because of the slight applied zero-point shift by the APOGEE calibration. We conclude that our fitted [$\alpha$/Fe] values match the APOGEE values within the error limits, and we are able to extract [$\alpha$/Fe] values very precisely (0.04\,dex average uncertainty) from the PO observed spectra.

The bottom right panel of Figure\,\ref{fig:synple_panels} displays the APOGEE$-$SONG [$\alpha$/Fe] as a function of SONG [$\alpha$/Fe], separately for the raw and the calibrated APOGEE values. The average $\Delta$[$\alpha$/Fe] is $-$0.03\,dex for the raw APOGEE values, and the standard deviation is 0.13\,dex. Due to the zero-point shift, the differences for the calibrated APOGEE parameters are slightly higher, but they are very close to the raw ones. In the case of calibrated parameters, the average $\Delta$[$\alpha$/Fe] is found to be $-$0.02\,dex, the standard deviation is 0.14\,dex, and the average error of SONG parameters is 0.07\,dex. Most $\Delta$[$\alpha$/Fe] are within the error limits for both raw and calibrated APOGEE parameters. We conclude that SONG [$\alpha$/Fe] values are in good agreement with APOGEE ones considering the uncertainties.

\section{Individual element abundances}\label{sec:abundance}

\begin{figure*}
    \begin{center}
      \includegraphics[width=0.9\textwidth]{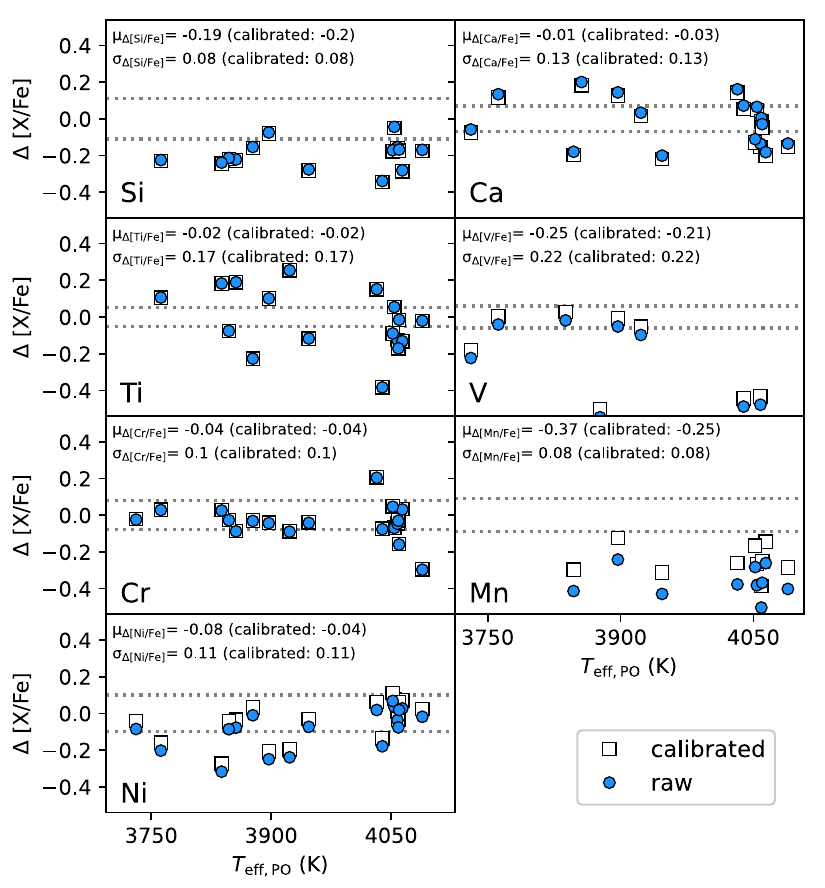}
    \end{center}
    \caption{[X/Fe] abundance differences between APOGEE and our fitted parameters as a function of $T_{\mathrm{eff}}$ values derived within this study. Blue filled circles and empty squares denote the raw and the calibrated APOGEE$-$PO values, respectively. The dotted grey lines represent the upper and lower average error limits. The average and the standard deviation of the differences are denoted by $\mu$ and $\sigma$ symbols, respectively.}
    \label{fig:po_abundance}
\end{figure*}

\begin{figure*}
    \begin{center}
      \includegraphics[width=0.9\textwidth]{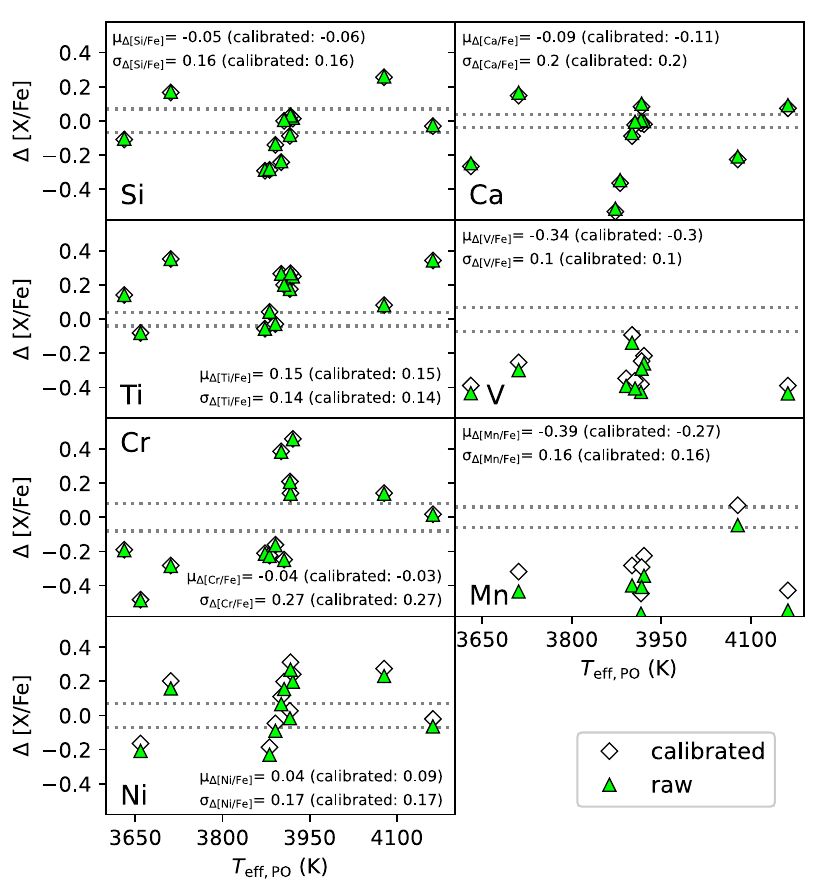}
    \end{center}
    \caption{Same as Figure\,\ref{fig:po_abundance} but for APOGEE$-$SONG abundances: green filled triangles and empty rhombi denote the raw and the calibrated values, respectively.}
    \label{fig:song_abundance}
\end{figure*}

In this section, we present and discuss the results of the analysis of Si, Ca, Ti, V, Cr, Mn, and Ni individual element abundances. Table\,\ref{tab:abundance} shows the results of the abundance analysis for our 21 targets. We compared our fitted abundances with the raw and the calibrated APOGEE values (if it is available). The raw APOGEE abundances are derived from atomic lines, the solar reference scale is based on \citet{Grevesseetal2007}. The calibrated APOGEE abundances have been shifted to force the solar-metallicity stars in the solar neighborhood to have [X/Fe] = 0 \citep{Jonssonetal2020}. The raw and the calibrated abundances are quite similar, therefore we do not treat them separately, and discuss only the results of raw values (aside from V, Ni, and Mn). Figures\,\ref{fig:po_abundance} and \ref{fig:song_abundance} show the [X/Fe] APOGEE$-$PO and APOGEE$-$SONG abundance differences as a function of our fitted $T_{\mathrm{eff}}$. The average [X/Fe] uncertainties are calculated from the quadratic mean error of the fitted parameters in this study and that of APOGEE.

\subsection{Silicon, Si}

The top left panel of Figure\,\ref{fig:po_abundance} shows the APOGEE$-$PO [Si/Fe] differences. Silicon is among the most precisely determined elements in APOGEE, but some APOGEE [Si/Fe] values may be super-solar for $T_{\mathrm{eff}}<4000$\,K. Based on the calculated NLTE models, Si\,I lines are not necessary to be corrected in the infrared APOGEE windows \citep{Jonssonetal2020}. We fitted 14 PO spectra in the window of the 5665.555\,\r{A} Si line and found that the average $\Delta$[Si/Fe] is $-$0.19\,dex. The $\Delta$[Si/Fe] standard deviation is found to be 0.08\,dex. While this scatter is considered small and shows that the PO measurements of Si\,I can be precise, though the average $-$0.19\,dex difference is larger than the 0.11\,dex average error range.

The APOGEE$-$SONG [Si/Fe] differences are shown in the top left panel of Figure\,\ref{fig:song_abundance}. We fitted 12 SONG spectra in the Si\,I window, as a result, the average and the standard deviation of $\Delta$[Si/Fe] are found to be $-$0.05\,dex and 0.16\,dex, respectively. In absolute terms, the average difference is better than the APOGEE$-$PO value. However, the scatter of the 12 APOGEE$-$SONG values is larger than that of the 14 APOGEE$-$PO abundances. Apart from the few low-abundance outlier SONG values (see Table\,\ref{tab:abundance}), the SONG Si\,I abundances are similar to the PO abundances. Overall, the Si\,I abundance differences can be considered slightly larger than expected in the case of APOGEE$-$SONG, too. This could indicate that the APOGEE Si\,I abundances have larger uncertainties for $T_{\mathrm{eff}}<4200$\,K than in the case of hotter stars.

\subsection{Calcium, Ca}

The APOGEE$-$PO [Ca/Fe] differences are presented in the top right panel of Figure\,\ref{fig:po_abundance}. Ca\,I has a high-precision APOGEE abundance with very few giants with $T_{\mathrm{eff}}<4000$\,K showing peculiar APOGEE abundance values in the super-solar metallicity range. The calibrated [Ca/Fe] values are based on the NLTE Ca\,I abundance calculations of \citet{Osorioetal2020}. 

Based on 16 fitted PO spectra, we found that the average $\Delta$[Ca/Fe] is $-$0.01\,dex in comparison with raw APOGEE abundances, and $-$0.03\,dex in comparison with calibrated values. The standard deviation of $\Delta$[Ca/Fe] is found to be 0.13\,dex both for raw and calibrated APOGEE values. We can conclude that the agreement between the APOGEE and PO [Ca/Fe] values is excellent.

The APOGEE$-$SONG [Ca/Fe] differences are shown in the top right panel of Figure\,\ref{fig:song_abundance}. Based on 11 fitted SONG spectra, the average $\Delta$[Ca/Fe] values are $-$0.09\,dex and $-$0.11\,dex for the raw and the calibrated APOGEE abundance, respectively. The standard deviation of $\Delta$[Ca/Fe] is found to be 0.2\,dex both for raw and calibrated values. We found a good agreement between the APOGEE and SONG Ca\,I abundances, however, the difference of the abundances is somewhat larger than the APOGEE$-$PO difference.

\subsection{Titanium, Ti}

The APOGEE$-$PO [Ti/Fe] differences can be seen in the 2$^{\mathrm{nd}}$ row left panel of Figure\,\ref{fig:po_abundance}. The Ti\,I abundance determined by APOGEE has moderate precision, however, the expected H-band [Ti/Fe] versus [Fe/H] trends significantly differ from the optical values. Thus, the reliable APOGEE data for the giants is available only for $T_{\mathrm{eff}} > 4200$\,K \citep{Jonssonetal2020}.  

The average $\Delta$[Ti/Fe] is $-$0.02\,dex based on 17 PO spectrum fittings. The standard deviation of $\Delta$[Ti/Fe] is found to be 0.17\,dex. Thus, our conclusion is that the PO [Ti/Fe] values closely match the APOGEE data considering the uncertainties. These new measurements may provide important reference values for APOGEE and may confirm that APOGEE Ti\,I abundances can be reliable for $T_{\mathrm{eff}} < 4200$\,K. 

The APOGEE$-$SONG [Ti/Fe] differences are shown in the 2$^{\mathrm{nd}}$ row left panel of Figure\,\ref{fig:song_abundance}. The average $\Delta$[Ti/Fe] is found to be 0.15\,dex, and the standard deviation is 0.14\,dex based on the reduction of 13 SONG spectra, which is larger than the mean APOGEE$-$PO Ti\,I abundance difference.

\begin{table*}
	\centering
    \fontsize{8pt}{12pt}\selectfont
	\caption{Element abundances of the stars observed with PO RCC and SONG telescopes.}
	\begin{tabular}{c r r r r r r r}
	\hline\hline

    \multirow{2}{*}{{2MASS \ ID}} & \multicolumn{7}{c}{PO} \\
    \cline{2-8}
	
    & \multicolumn{1}{c}{[Si/Fe]} & \multicolumn{1}{c}{[Ca/Fe]} & \multicolumn{1}{c}{[Ti/Fe]} & \multicolumn{1}{c}{[V/Fe]} & \multicolumn{1}{c}{[Cr/Fe]} & \multicolumn{1}{c}{[Mn/Fe]} & \multicolumn{1}{c}{[Ni/Fe]}\\
	\hline
    J01094391+3537137   & 0.25 $\pm$ 0.10 & $-$0.18 $\pm$ 0.06 &    0.10 $\pm$ 0.04 & $-$0.13 $\pm$ 0.04 & $-$0.03 $\pm$ 0.07 & $\cdots$            &    0.17 $\pm$ 0.10 \\
    J01301114+0608377   & 0.22 $\pm$ 0.15 & $-$0.02 $\pm$ 0.08 &    0.10 $\pm$ 0.06 &  $\cdots$            &    0.00 $\pm$ 0.10 &   0.20 $\pm$ 0.10 & $-$0.01 $\pm$ 0.12 \\
    J04200995+3157113   & $\cdots$          & $-$0.08 $\pm$ 0.07 & $-$0.06 $\pm$ 0.05 & $-$0.08 $\pm$ 0.05 &    0.05 $\pm$ 0.06 & $\cdots$            &    0.18 $\pm$ 0.09 \\
    J05022869+4104329   & 0.30 $\pm$ 0.12 & $-$0.20 $\pm$ 0.06 &    0.32 $\pm$ 0.06 &    0.24 $\pm$ 0.05 &    0.06 $\pm$ 0.07 & $\cdots$            &    0.14 $\pm$ 0.11 \\
    J05062972+6110109   & 0.22 $\pm$ 0.15 &    0.06 $\pm$ 0.08 &    0.10 $\pm$ 0.06 &  $\cdots$            &    0.03 $\pm$ 0.10 &   0.26 $\pm$ 0.10 & $-$0.05 $\pm$ 0.12 \\
    J06300297+4641079   & 0.17 $\pm$ 0.14 &    0.09 $\pm$ 0.07 &    0.09 $\pm$ 0.06 &  $\cdots$            & $-$0.03 $\pm$ 0.09 &   0.20 $\pm$ 0.09 & $-$0.10 $\pm$ 0.11 \\
    J07363163+4610488   & 0.23 $\pm$ 0.06 &  $\cdots$            &    0.16 $\pm$ 0.04 & $-$0.16 $\pm$ 0.04 & $-$0.01 $\pm$ 0.05 & $\cdots$            &    0.32 $\pm$ 0.07 \\
    J07405852+2301067   & 0.13 $\pm$ 0.07 & $-$0.27 $\pm$ 0.05 &    0.11 $\pm$ 0.05 & $-$0.04 $\pm$ 0.05 &    0.05 $\pm$ 0.06 &   0.33 $\pm$ 0.08 &    0.28 $\pm$ 0.09 \\
    J08555556+1137335   & 0.14 $\pm$ 0.13 &    0.10 $\pm$ 0.06 & $-$0.05 $\pm$ 0.05 &  $\cdots$            &    0.08 $\pm$ 0.08 &   0.39 $\pm$ 0.09 & $-$0.02 $\pm$ 0.11 \\
    J09413511+3116398   & $\cdots$          &    0.02 $\pm$ 0.07 &  $\cdots$            &    0.05 $\pm$ 0.04 &    0.05 $\pm$ 0.07 & $\cdots$            &    0.06 $\pm$ 0.09 \\
    J10254427$-$0703358 & 0.23 $\pm$ 0.07 & $-$0.21 $\pm$ 0.06 &    0.06 $\pm$ 0.05 &  $\cdots$            &    0.10 $\pm$ 0.06 & $\cdots$            &    0.01 $\pm$ 0.06 \\
    J11301888$-$0300128 & 0.21 $\pm$ 0.12 &    0.16 $\pm$ 0.07 &    0.14 $\pm$ 0.05 &  $\cdots$            &    0.00 $\pm$ 0.08 &   0.30 $\pm$ 0.09 &    0.02 $\pm$ 0.10 \\
    J13363360+5241148   & 0.14 $\pm$ 0.07 &  $\cdots$            &    0.21 $\pm$ 0.05 &    0.26 $\pm$ 0.04 &    0.08 $\pm$ 0.05 & $\cdots$            & $-$0.01 $\pm$ 0.07 \\
    J14490671+5413537   & $\cdots$          & $-$0.04 $\pm$ 0.08 &    0.03 $\pm$ 0.07 &  $\cdots$            & $-$0.19 $\pm$ 0.08 &   0.20 $\pm$ 0.10 &    0.00 $\pm$ 0.11 \\
    J15261738+3420095   & $\cdots$          &    0.01 $\pm$ 0.08 &    0.24 $\pm$ 0.06 &  $\cdots$            &    0.05 $\pm$ 0.10 &   0.32 $\pm$ 0.10 &    0.02 $\pm$ 0.12 \\
    J16131544+0501160   & 0.25 $\pm$ 0.13 &    0.11 $\pm$ 0.07 &    0.24 $\pm$ 0.07 &  $\cdots$            & $-$0.05 $\pm$ 0.07 &   0.23 $\pm$ 0.09 & $-$0.07 $\pm$ 0.10 \\
    J17214533+5325135   & 0.23 $\pm$ 0.13 &    0.15 $\pm$ 0.07 &    0.18 $\pm$ 0.06 &  $\cdots$            & $-$0.02 $\pm$ 0.08 &   0.34 $\pm$ 0.09 & $-$0.03 $\pm$ 0.11 \\
    J19363755+4830583   & 0.15 $\pm$ 0.12 &    0.08 $\pm$ 0.07 &    0.18 $\pm$ 0.05 &    0.10 $\pm$ 0.05 &    0.07 $\pm$ 0.07 & $\cdots$            &    0.03 $\pm$ 0.11 \\
    \hline
	
    \multirow{2}{*}{{2MASS ID}} & \multicolumn{7}{c}{SONG} \\
    \cline{2-8}
    
    & \multicolumn{1}{c}{[Si/Fe]} & \multicolumn{1}{c}{[Ca/Fe]} & \multicolumn{1}{c}{[Ti/Fe]} & \multicolumn{1}{c}{[V/Fe]} & \multicolumn{1}{c}{[Cr/Fe]} & \multicolumn{1}{c}{[Mn/Fe]} & \multicolumn{1}{c}{[Ni/Fe]}\\
    \hline
    J01301114+0608377   &    0.17 $\pm$ 0.07 &    0.05 $\pm$ 0.04 & $-$0.05 $\pm$ 0.04 &    0.24 $\pm$ 0.05 &    0.18 $\pm$ 0.08 & $\cdots$              & $-$0.13 $\pm$ 0.09 \\
    J04200995+3157113   &    0.26 $\pm$ 0.06 &    0.30 $\pm$ 0.03 &    0.16 $\pm$ 0.04 &  $\cdots$            &    0.19 $\pm$ 0.06 & $\cdots$            &    0.17 $\pm$ 0.06 \\
    J05022869+4104329   & $-$0.30 $\pm$ 0.06 &    0.08 $\pm$ 0.06 & $-$0.15 $\pm$ 0.06 &  $\cdots$            & $-$0.16 $\pm$ 0.09 &   0.10 $\pm$ 0.06 & $-$0.27 $\pm$ 0.06 \\
    J05062972+6110109   &    0.02 $\pm$ 0.07 & $-$0.07 $\pm$ 0.05 & $-$0.18 $\pm$ 0.04 &    0.07 $\pm$ 0.06 & $-$0.27 $\pm$ 0.08 &   0.30 $\pm$ 0.06 & $-$0.30 $\pm$ 0.08 \\
    J05544363$-$1146270 & $-$0.04 $\pm$ 0.08 & $-$0.04 $\pm$ 0.04 & $-$0.14 $\pm$ 0.04 &    0.30 $\pm$ 0.04 &    0.29 $\pm$ 0.07 &   0.30 $\pm$ 0.05 & $-$0.18 $\pm$ 0.06 \\
    J06300297+4641079   & $-$0.02 $\pm$ 0.06 & $-$0.02 $\pm$ 0.04 & $-$0.25 $\pm$ 0.04 & $-$0.01 $\pm$ 0.04 & $-$0.44 $\pm$ 0.06 &   0.26 $\pm$ 0.07 & $-$0.23 $\pm$ 0.08 \\
    J07363163+4610488   &  $\cdots$            &  $\cdots$            &    0.42 $\pm$ 0.03 &  $\cdots$            &    0.50 $\pm$ 0.06 & $\cdots$       &    0.21 $\pm$ 0.04 \\
    J07405852+2301067   &    0.19 $\pm$ 0.05 &  $\cdots$            &    0.24 $\pm$ 0.04 &    0.30 $\pm$ 0.05 &    0.17 $\pm$ 0.05 & $\cdots$      &    0.12 $\pm$ 0.05 \\
    J09413511+3116398   &    0.11 $\pm$ 0.07 &    0.21 $\pm$ 0.03 &    0.00 $\pm$ 0.04 &    0.26 $\pm$ 0.03 &    0.22 $\pm$ 0.07 & $\cdots$       &  $\cdots$            \\
    J14153968+1910558   &    0.22 $\pm$ 0.07 & $-$0.00 $\pm$ 0.05 & $-$0.24 $\pm$ 0.06 &    0.21 $\pm$ 0.05 & $-$0.08 $\pm$ 0.10 &   0.32 $\pm$ 0.05 &    0.12 $\pm$ 0.10 \\
    J15261738+3420095   &    0.14 $\pm$ 0.07 &    0.01 $\pm$ 0.04 & $-$0.11 $\pm$ 0.04 &    0.14 $\pm$ 0.06 & $-$0.19 $\pm$ 0.08 &   0.38 $\pm$ 0.06 & $-$0.04 $\pm$ 0.09 \\
    J17214533+5325135   &    0.19 $\pm$ 0.07 &    0.02 $\pm$ 0.04 & $-$0.20 $\pm$ 0.03 & $-$0.18 $\pm$ 0.04 & $-$0.45 $\pm$ 0.05 &   0.31 $\pm$ 0.06 & $-$0.17 $\pm$ 0.09 \\
    J22290798+0907446   &    0.26 $\pm$ 0.06 &    0.45 $\pm$ 0.05 &    0.19 $\pm$ 0.04 &  $\cdots$            &    0.18 $\pm$ 0.07 & $\cdots$        &  $\cdots$            \\
	\hline
	
	\end{tabular}
	\label{tab:abundance}
\end{table*}

\subsection{Vanadium, V}

The APOGEE$-$PO differences of the V\,I abundance are shown in the 2$^{\mathrm{nd}}$ row right panel of Figure\,\ref{fig:po_abundance}. V\,I is one of the elements which have the least precisely determined APOGEE abundance \citep{Jonssonetal2020}. Also, PO abundances are available only for 8 stars because of the weak V\,I line in every spectra. 

We found that the average $\Delta$[V/Fe] for the raw APOGEE data is $-$0.25\,dex with 0.22\,dex of standard deviation. For the calibrated APOGEE abundance, the average is found to be $-$0.21\,dex, and the standard deviation is 0.22\,dex. The calibrated $\Delta$[V/Fe] values are slightly larger than the raw values because of the zero-point shift. Therefore, the PO V\,I  matching is better with the calibrated APOGEE values than with the raw values.

A systematic negative offset can be observed in $\Delta$[V/Fe] for the eight PO targets, which is mostly caused by three outliers with peculiarly high [V/Fe] PO abundance, though there are four stars in our sample whose [V/Fe] values agree very well with APOGEE. Based on these findings our conclusion is that fitting the 4594.08\,\r{A} V line in our PO observations is more difficult than Ca, Ti, or Cr and small zero-point offsets may exist in our data.

The APOGEE$-$SONG difference of V\,I abundance are shown in the 2$^{\mathrm{nd}}$ row right panel of Figure\,\ref{fig:song_abundance}. Based on nine fitted SONG spectra, the average $\Delta$[V/Fe] is found to be $-$0.34\,dex, and the standard deviation is 0.1\,dex. One can see that the SONG [V/Fe] values systematically overshoot the APOGEE abundances, confirming the results of the PO observations.

\subsection{Chromium, Cr}

The 3$^{\mathrm{rd}}$ row left panel of Figure\,\ref{fig:po_abundance} presents the APOGEE$-$PO differences of the abundance of Cr\,I. The [Cr/Fe] is determined with medium precision and accuracy in APOGEE data \citep{Jonssonetal2020}. 

We determined the Cr\,I abundances for all of the 18 PO spectra. We found that the average $\Delta$[Cr/Fe] is $-$0.04\,dex, and the standard deviation of $\Delta$[Cr/Fe] is 0.1\,dex. The excellent agreement with APOGEE and the relatively small scatter of the differences indicates the PO Cr\,I is one of the most precise elements we can measure and derive as well. Our measurements are considered to be high-quality, which is reinforced by the fact that three Cr lines are available to fit in PO spectra (the average uncertainty of PO [Cr/Fe] is 0.08\,dex).

The APOGEE$-$SONG differences of Cr\,I abundance are shown in the 3$^{\mathrm{rd}}$ row left panel of Figure\,\ref{fig:song_abundance}. We fitted 13 SONG spectra in three Cr\,I windows, and found that the average $\Delta$[Cr/Fe] is $-$0.04\,dex, and the standard deviation is 0.27\,dex. Cr\,I abundances of 3 stars show good agreement with APOGEE data, however, the scatter of Cr\,I abundances is relatively large. The average is the same as the APOGEE$-$PO value, while the scatter is definitely larger than the that of APOGEE$-$PO.

\subsection{Manganese, Mn}

The 3$^{\mathrm{rd}}$ row right panel of Figure\,\ref{fig:po_abundance} shows APOGEE$-$PO differences of the abundance of Mn\,I. The [Mn/Fe] is determined with high precision in APOGEE DR16 \citep{Jonssonetal2020}. However, it was calibrated with a relatively large zero-point offset for giants, and the database is mainly populated only for $T_{\mathrm{eff}}>4000$\,K giants. We note that our line list contains hyper-fine splitting for Mn.

We determined the Mn\,I abundance of 10 PO targets. The average and the standard deviation of $\Delta$[Mn/Fe] are $-$0.37, and 0.08\,dex, respectively, for the raw APOGEE parameters. The relatively low scatter shows that PO measurements are consistent for the Mn\,I abundance. The $-$0.37\,dex average difference is larger than the 0.09\,dex mean error range, which can be traced back to a systematic overestimation of the Mn\,I abundances. The average and the standard deviation of the calibrated $\Delta$[Mn/Fe] are $-$0.25 and 0.08\,dex. The scatter is still considered small, however, the mean difference between the calibrated APOGEE and the PO values is smaller than between the raw APOGEE and the PO. 

The APOGEE$-$SONG differences of Mn\,I abundance are shown in the 3$^{\mathrm{rd}}$ row right panel of Figure\,\ref{fig:song_abundance}. Only 7 SONG target's Mn\,I abundances were determined because of the weak 5420.425\,\,\r{A} Mn line. The average $\Delta$[Mn/Fe] is found to be $–$0.39\,dex, and the standard deviation is 0.16\,dex for the raw APOGEE data. Similarly to PO Mn\,I abundances, the SONG [Mn/Fe] values systematically overestimate the APOGEE data. The average of the calibrated $\Delta$[Mn/Fe] is $–$0.27\,dex, and the standard deviation is 0.16\,dex. While the SONG abundances are found to be closer to the calibrated APOGEE values than that of the raw values, even so, the APOGEE$-$SONG Mn\,I abundances are generally out of the 0.06\,dex error range. 

We conclude that the SONG Mn\,I abundances confirm the PO [Mn/Fe] values. However, the Mn\,I abundances of both PO and SONG observations systematically overestimate the APOGEE data. While our testing shows that the selected Mn line is sensitive to the abundance of Mn\,I in this temperature, the sensitivity is weak which might cause the large systematic offset seen in the comparison with APOGEE data. The lack of NLTE corrections in the optical wavelength range could explain the systematic offset from the H-band abundance values of APOGEE.

\subsection{Nickel, Ni}

The bottom left panel of Figure\,\ref{fig:po_abundance} presents the APOGEE$-$PO differences of the Ni\,I abundance. Ni\,I is one of the most precisely determined elements in APOGEE \citep{Jonssonetal2020}, but with a zero-point offset slightly larger than for the rest of the elements discussed in this section. 

Here, the Ni\,I abundances were determined for 18 PO spectra. We found that the average $\Delta$[Ni/Fe] for the raw and calibrated APOGEE abundances are $-$0.08 and $-$0.04\,dex, respectively. The standard deviation of $\Delta$[Ni/Fe] is 0.11\,dex both for raw and calibrated APOGEE abundances. The PO Ni\,I abundances show good agreement with APOGEE, the low scatter of the differences indicates that the PO [Ni/Fe] values are precise, similarly to Cr.

The bottom left panel of Figure\,\ref{fig:song_abundance} presents the APOGEE$-$SONG differences of Ni\,I abundance. The average $\Delta$[Ni/Fe] values are 0.04\,dex and 0.09\,dex for the raw and the calibrated APOGEE data, respectively, based on 11 SONG spectra. The standard deviation of $\Delta$[Ni/Fe] is 0.17\,dex both for raw and calibrated APOGEE data. The SONG observations confirm PO very well. Although, the SONG Ni\,I abundances match better with the raw APOGEE data, while the PO abundances are closer to the calibrated APOGEE values. 

Considering that the average raw $\Delta$[Ni/Fe] of SONG is inside the 0.07\,dex error range, and also the average $\Delta$[Ni/Fe] of PO (for both raw and calibrated APOGEE data) is inside the 0.1\,dex error range, we conclude that the agreements between the APOGEE and SONG, as well as APOGEE and PO, are excellent.

\section{Summary}\label{sec:conclusions}

We obtained high-resolution spectra of 21 red giant stars with the spectrograph mounted on the Piszk\'estet\H{o} Observatory and the SONG 1-m telescope and determined their atmospheric parameters and abundances of seven elements. We took advantage of 
the new BOSZ synthetic spectral library originally developed for the flux calibration of the James Webb telescope (M\'esz\'aros et al. in prep). This new library was calculated using the line lists of 23 molecules that make it possible to properly model the observed spectra of our targets.

We found that the average $T_{\mathrm{eff}}$ difference between the APOGEE and Piszk\'estet\H{o} observations is $-$11.2\,K. The average $\log g$ difference is found to be 0.11\,dex, and the average [Fe/H] and [$\alpha$/Fe] differences are 0.1 and $-$0.01\,dex, respectively. Considering our $\log g$, and [$\alpha$/Fe] uncertainties we conclude that our determined parameters generally agree with that of the APOGEE. Our fitted individual element abundances also match well the APOGEE values, especially in the case of Si, Ca, Ti, Cr, and Ni. We could not completely reproduce all the APOGEE values for Mn, and V abundances, we note that even the APOGEE recommendation is to handle these elements with caution \citep{Jonssonetal2020}. 

This is the first time that the high-resolution spectrographs mounted on the 1-m telescopes at Piszk\'estet\H{o} and the SONG Observatory were used for abundance analysis. We showed that it is possible to measure $T_{\mathrm{eff}}$ within 50\,K for $T_{\mathrm{eff}}<4200$\,K, and $\log g$ with 0.2\,dex precision between $\log g = 1$ and 2 in the 4500$-$5800\,\r{A} wavelength range from Piszk\'estet\H{o}. The 0.1\,dex precision can be generally achieved in the abundance measurements. The average temperature difference between APOGEE and SONG observations is 82.3\,K, the average $\log g$ difference is $-$0.09\,dex, and the average abundance differences vary between 0.23 and $-$0.03\,dex which is close to the estimated uncertainties of both measurements.

We successfully demonstrated that in the case of relatively bright ($G<8-9$\,mag) red giant stars reliable atmospheric parameters and element abundances can be derived with the high-resolution spectrographs mounted on the 1-m Piszk\'estet\H{o} and the SONG observatories. On the other hand, our results show that observing programs similar to ours with relatively small telescopes can be used to accommodate the observations of large spectroscopic sky surveys like SDSS/APOGEE.

\begin{acknowledgments}

We would like to thank our referee whose comments were invaluable to us when preparing this paper. This project has been supported by the LP2021-9 Lend\"ulet grant of the Hungarian Academy of Sciences. This work includes observations made with the Hertzsprung SONG Telescope operated at the Spanish Observatorio del Teide on the island of Tenerife by the Aarhus University and by the Instituto de Astrofísica de Canarias. B.Cs. acknowledges support from the Lend\"ulet Program LP2023-10 of the Hungarian Academy of Sciences. On behalf of the "Calculating the Synthetic Stellar Spectrum Database of the James Webb Space Telescope" project we are grateful for the possibility of using HUN-REN Cloud (see Héder et al. 2022; https://science-cloud.hu/) which helped us achieve the results published in this paper. We acknowledge the funding of the National Research, Development and Innovation Office of Hungary (NKFIH) via the grant K-138962. V.H. is supported by the DKOP-23 Doctoral Excellence Program of the Ministry for Culture and Innovation from the source of the National Research, Development and Innovation Fund. This work was supported by the PRODEX Experiment Agreement No. 4000137122 between the ELTE E\"otv\"os Lor\'and University and the European Space Agency (ESA-D/SCI-LE-2021-0025). This project has received funding from the HUN-REN Hungarian Research Network.
      
\end{acknowledgments}

\bibliography{main_pasp.bib}{}
\bibliographystyle{aasjournal}

\end{document}